# Determination of Burgers vectors of dislocations in monoclinic β-$Ga_2O_3$ crystals by large-angle convergent-beam electron diffraction

Yoshihiro Sugawara [1,*], Yongzhao Yao [1, 2], and Yukari Ishikawa [1]

[1]Japan Fine Ceramics Center, 2-4-1 Mutsuno, Atsuta, Nagoya 456-8587, Japan

[2]Innovation Center for Semiconductor and Digital Future, Mie University, 1577, Kurimamachiya-cho, Tsu 514-8507, Japan

We demonstrate the applicability of large-angle convergent-beam electron diffraction (LACBED) for Burgers vector determination in monoclinic β-$Ga_2O_3$. The inner product $\boldsymbol{g}\cdot\boldsymbol{b}$ in this non-orthogonal system can be evaluated without a metric tensor by using the dual relationship between real and reciprocal lattice bases. Based on this framework, Burgers vectors of dislocations introduced by nanoindentation were unambiguously determined from LACBED node counts. The results are consistent with weak-beam dark-field imaging, confirming the effectiveness of LACBED for β-$Ga_2O_3$.

* Author to whom correspondence should be addressed. Electronic mail: sugawara@jfcc.or.jp

$\beta$-$Ga_2O_3$ has recently attracted significant attention as a promising semiconductor for realizing high-efficiency and ultra-high-voltage power conversion devices.[1,2,3] It possesses attractive material properties compared with other semiconductors, such as a wide bandgap of approximately 4.5 eV and a high breakdown electric field of about 7 MV/cm.[1,2] Large-diameter $\beta$-$Ga_2O_3$ single-crystal bulk can be grown by various melt-growth techniques,[1,4,5,6] and $\beta$-$Ga_2O_3$ substrates with diameters of up to 4 inches fabricated by the edge-defined film-fed growth (EFG) method are now commercially available.[1]

However, a large number of dislocations still remain in the crystal,[7] and their impact on device performance is expected.[8,9,10] Therefore, it is important to gain a deep understanding of dislocation structures and to establish process control guidelines for reducing dislocations. In SiC and GaN, it is essential not only to determine the direction of the Burgers vector of dislocations but also to identify its sense and magnitude in order to clarify their effects on device characteristics. For example, screw dislocations in SiC are closed-core up to 2c, but are known to become hollow micropipes when the Burgers vector exceeds 3c,[11] which act as major leakage paths.[12,13] Basal plane dislocations (BPDs) that expand into stacking faults and lead to forward degradation[14] exhibit a slight dependence of the ease of benign conversion via BPD-to-threading edge dislocation transformation on the sense of their Burgers vectors.[15] Similar phenomena are also expected in β-$Ga_2O_3$, and accurate determination of the Burgers vectors of dislocations is therefore required.

In general, weak-beam dark-field (WBDF) imaging is used for dislocation structure analysis.[16] However, since WBDF is an imaging technique based on the relation $\boldsymbol{g} \cdot \boldsymbol{b} = 0$ (where $\boldsymbol{g}$ is the reciprocal lattice vector and $\boldsymbol{b}$ is the Burgers vector), it allows determination of the direction of the Burgers vector $\boldsymbol{b}$, but it is difficult to determine its magnitude. In such cases, large-angle convergent-beam electron diffraction (LACBED), which enables unique determination of the Burgers vector using the relation $\boldsymbol{g} \cdot \boldsymbol{b} = n$ (where $n$ is the number of nodes of the diffraction line), is effective.[17,18,19,20,21] The usefulness of this method has already been demonstrated in cubic Si and hexagonal materials such as SiC[22] and GaN.[23,24,25,26] On the other hand, β-$Ga_2O_3$ has a monoclinic crystal structure, and its applicability has not yet been sufficiently investigated. Therefore, it is important to demonstrate that LACBED can also be applied to β-$Ga_2O_3$.

In this study, we examined the treatment of vector inner products in non-orthogonal crystal systems such as β-$Ga_2O_3$. As a model case, dislocations were introduced by nanoindentation[27] using a Berkovich indenter,[28] and subsequently measured and analyzed by LACBED. The validity of this approach was evaluated by comparing the obtained results with those from WBDF imaging.

A $(\bar{2}01)$-oriented single-crystal wafer with a thickness of 680 μm, grown by the edge-defined film-fed growth (EFG) method,[4] was used in this study. The wafer was doped with Sn to achieve a donor concentration of $4.6 \times 10^{18}$ $cm^{-3}$. X-ray diffraction (XRD) measurements showed that the full width at half maximum (FWHM) was 20 arcsec along the [010] azimuth and 17 arcsec along the [102] azimuth, indicating high crystalline quality of the wafer. The dislocation density in the wafer was estimated to be on the order of $10^4$ $cm^{-2}$ using X-ray topography. The surface was polished by chemical mechanical polishing (CMP).

Nanoindentation[27] was performed using a Berkovich indenter (a three-sided pyramidal diamond tip) with a Bruker Hysitron TS77 system. The maximum load was 50 mN, and the test temperature was maintained at 25 ± 0.5 °C. The β-$Ga_2O_3$ $(\bar{2}01)$ wafer was oriented such that one facet of the pyramidal indenter was aligned parallel to the [102] direction of β-$Ga_2O_3$.[28] During the indentation experiment, it took 5 sec for the indenter to reach the peak load after initial contact with the sample surface. The peak load was then held for 2 sec, followed by unloading over 5 sec to withdraw the indenter.

A cross-sectional TEM specimen was prepared by focused ion beam (FIB) microsampling[29,30] along a plane perpendicular to the [102] direction and passing through the center of the indentation. A cross-sectional sample with dimensions of 14 μm × 10 μm × 150 nm was fabricated so as to include the indentation and the surrounding damaged region. FIB processing was carried out using a Hitachi High-Tech dual-beam FIB system (NB5000). The dislocation structure was characterized by LACBED[17,18,19,20,21] and WBDF[16] methods. Dislocation observations were performed using a JEOL JEM-2100Plus transmission electron microscope operated at an accelerating voltage of 200 kV. LACBED patterns were acquired near the [102] zone axis. WBDF images were obtained under two-beam conditions with reciprocal lattice vectors $\boldsymbol{g} = \bar{2}\ 0\ 1$ and $\boldsymbol{g} = 0\ 2\ 0$. CBED pattern analysis was performed using Ideal Microscope.

In this section, the determination of Burgers vectors using the relation $\boldsymbol{g}\cdot\boldsymbol{b} = n$ is described for non-orthogonal crystal systems such as β-$Ga_2O_3$.

In general, when calculating the inner product of two vectors $\boldsymbol{v}$ and $\boldsymbol{w}$ in a non-orthogonal coordinate system, a metric tensor $G$ is required, as expressed by Eq. (1):

$$\boldsymbol{v} \cdot \boldsymbol{w} = v^i G_{ij} w^j. \tag{1}$$

However, real-space (direct) and reciprocal-space bases are defined such that a dual (biorthogonal) relationship is satisfied. That is, even in non-orthogonal crystal systems, the use of dual bases enables the calculation of inner products without explicitly introducing the metric tensor.

The three lattice vectors of a non-orthogonal crystal system are denoted by $\boldsymbol{a}_1$, $\boldsymbol{a}_2$, and $\boldsymbol{a}_3$, and the real-space basis is given by the set of these three linearly independent vectors as

$$\{\boldsymbol{a}_j\}, \quad j = 1, 2, 3. \tag{2}$$

Correspondingly, the reciprocal-space basis vectors $\boldsymbol{g}_1$, $\boldsymbol{g}_2$, and $\boldsymbol{g}_3$ are expressed as

$$\{\boldsymbol{g}_i\}, \quad i = 1, 2, 3. \tag{3}$$

As mentioned above, a dual relationship between the real-space and reciprocal-space bases is defined irrespective of whether the coordinate system is orthogonal or non-orthogonal, as given by Eq. (4):

$$\boldsymbol{g}_i \cdot \boldsymbol{a}_j = 2\pi\delta_{ij}. \tag{4}$$

Here, $\delta_{ij}$is the Kronecker delta, which takes a value of $2\pi$ when $i = j$ and 0 when $i \neq j$. In the present formulation, the factor of $2\pi$ is included in the definition of the dual basis to ensure invariance of the wave function $e^{i\boldsymbol{k}\cdot\boldsymbol{r}}$under lattice translation, i.e., $\boldsymbol{k} \cdot \boldsymbol{a} = 2\pi n$, where $\boldsymbol{k}$ is the wave vector, $\boldsymbol{r}$ is the position vector in real space, and $\boldsymbol{a}$ is a lattice translation vector. Thus, even in non-orthogonal crystal systems, the reciprocal basis vectors $\boldsymbol{g}_i$ are normalized with a factor of $2\pi$ so as to form an orthonormal dual basis with respect to $\boldsymbol{a}_j$.

When a dislocation exists in a non-orthogonal crystal system, an arbitrary Burgers vector $\boldsymbol{b}$ can be expressed using integer coefficients $n_j$ as

$$\boldsymbol{b} = n_1\boldsymbol{a}_1 + n_2\boldsymbol{a}_2 + n_3\boldsymbol{a}_3, \qquad n_j \in \mathbb{Z}. \tag{5}$$

Similarly, an arbitrary reciprocal lattice vector $\boldsymbol{g}$ can be written as a linear combination of the reciprocal basis vectors with integer coefficients $h_i$ as

$$\boldsymbol{g} = h_1\boldsymbol{g}_1 + h_2\boldsymbol{g}_2 + h_3\boldsymbol{g}_3, \qquad h_i \in \mathbb{Z}. \tag{6}$$

Here, we evaluate $\boldsymbol{g} \cdot \boldsymbol{b}$:

$$\boldsymbol{g} \cdot \boldsymbol{b} = \left(\sum_i h_i\, \boldsymbol{g}_i\right) \cdot \left(\sum_j n_j\, \boldsymbol{a}_j\right) = \sum_{i,j} h_i\, n_j\left(\boldsymbol{g}_i \cdot \boldsymbol{a}_j\right). \tag{7}$$

By applying the definition of the dual basis in Eq. (4), we obtain

$$\boldsymbol{g} \cdot \boldsymbol{b} = \sum_{i,j} h_i\, n_j\left(2\pi\delta_{ij}\right) = 2\pi \sum_i h_i\, n_i = 2\pi N, \quad N \in \mathbb{Z}. \tag{8}$$

Thus, even in non-orthogonal crystal systems, the inner product between real-space and reciprocal-space basis vectors can be evaluated straightforwardly without using the metric tensor, owing to the dual-basis relationship. Consequently, the $\boldsymbol{g} \cdot \boldsymbol{b}$ product can be calculated in a simple manner.

Fig. 1 shows a cross-sectional bright-field scanning transmission electron microscopy (BF-STEM) image of dislocations introduced around an indentation by nanoindentation. An indentation is formed near the center of the bright-field image. Beneath the indentation, dislocations and their associated strain fields are observed as dark contrast. In particular, a dislocation-rich region is present directly beneath the indentation; however, in this region, the strain fields of individual dislocations overlap, making it difficult to analyze each dislocation separately. Therefore, in this study, dislocations D-1 to D-8 located in the surrounding region, where the influence of the strain field is relatively small, were selected for analysis, and CBED patterns were acquired for each of them.

Fig. 2 shows three bright-field LACBED patterns obtained from dislocation D-1. In the figures, the white dashed line D indicates the dislocation line, and L, marked between white arrows, denotes the Laue reflection line. At the intersection between the dislocation line and the Laue reflection line in the LACBED patterns, $n$ nodes corresponding to the relation $\boldsymbol{g} \cdot \boldsymbol{b} = n$ (with $\boldsymbol{b} = [u, v, w]$) are observed. Figs 2(a), (b), and (c) correspond to the reciprocal lattice vectors $\boldsymbol{g}_1 = \bar{2}\,\bar{2}\,1$, $\boldsymbol{g}_2 = 7\,3\,\bar{3}$, and $\boldsymbol{g}_3 = \bar{8}\,\bar{2}\,4$, respectively. The numbers of nodes are $n_1 = 2$, $n_2 = -3$, and $n_3 = -2$, respectively. The sign of $n$ was determined based on the Cherns–Preston rule from the direction of the excitation error $s$.[19] The nodes appearing at the intersection positions are indicated by yellow arrows in the figures, and their numbers correspond to the respective values of $n$. From Figs. 2(a)–(c), the following equations are obtained:

$$-2u - 2v + w = 2 \quad (9)$$

$$7u + 3v - 3w = -3 \quad (10)$$

$$-8u - 2v + 4w = -2 \quad (11)$$

By solving this system of three linear equations, the Burgers vector $\boldsymbol{b}$ is determined to be $\boldsymbol{b} = [0\ \bar{1}\ 0]$. Furthermore, LACBED analysis was performed on dislocations D-2 to D-8 using the same method, and the resulting Burgers vectors are summarized in Table 1. As a result of the analysis, all of these dislocations were found to have a Burgers vector of $\boldsymbol{b} = \langle 010 \rangle$. The reason why all dislocations around the indentation possess the same Burgers vector is beyond the scope of this study, and therefore a detailed discussion is not provided here.

To verify the Burgers vectors determined above, a correlation was made with the dislocation contrast obtained from WBDF imaging. Fig. 3 shows WBDF images of dislocations formed around an indentation. The two-beam excitation conditions used for observation were $\boldsymbol{g} = \bar{2}\ 0\ 1$ for Fig. 3(a) and $\boldsymbol{g} = 0\ 2\ 0$ for Fig. 3(b), both under the $\boldsymbol{g}/3\boldsymbol{g}$ weak-beam condition. Dislocation D-1 is located within the yellow square indicated in the figures, and its magnified images are shown in Figs. 3(c) and (d). In Fig. 3(c), the dislocation contrast is very weak, whereas in Fig. 3(d), a clear contrast is observed. Thus, the Burgers vector of D-1 is parallel to [010] direction. Consistently, the dislocation contrasts of D-1 to D-8 are all very weak for the reciprocal lattice vector $\boldsymbol{g} = \bar{2}\ 0\ 1$, while they are clearly observed for $\boldsymbol{g} = 0\ 2\ 0$. These observations are consistent with the result obtained by LACBED that the Burgers vectors of dislocations D-1 to D-8 are $\boldsymbol{b} = \langle 0\ 1\ 0 \rangle$, thereby supporting the validity of the LACBED analysis.

In conclusion, in the analysis of dislocation structures in β-$Ga_2O_3$ (a monoclinic, non-orthogonal crystal system), it was demonstrated that the inner product of vectors can be evaluated straightforwardly without using a metric tensor, as in orthogonal crystal systems, indicating that the LACBED method is applicable. Model dislocations introduced into β-$Ga_2O_3$ by nanoindentation were analyzed by LACBED, and their Burgers vectors were determined. The results were consistent with those obtained by WBDF imaging. These findings demonstrate that the LACBED method is effective for determining the Burgers vectors of dislocations in β-$Ga_2O_3$ and is expected to contribute to a better understanding of defects affecting device performance.

**Acknowledgments**

Part of this work was supported by a project commissioned by the New Energy and Industrial Technology Development Organization (NEDO, JPNP22007) and by Japan Society for the Promotion of Science KAKENHI (Grant No. 25K08499). The authors gratefully acknowledge Dr. K. Sasaki at Novel Crystal Technology Inc. for providing the β-$Ga_2O_3$ substrates.

---

## Table Caption List

Table 1. Burgers vectors of various types of dislocations determined from bright-field LACBED patterns. The Burgers vectors were unambiguously determined from the number of nodes observed in the LACBED patterns.

## Figure Captions List

Fig. 1. Cross-sectional BF-STEM image of dislocations introduced around an indentation in β-$Ga_2O_3$ by nanoindentation. The substrate surface is the $(\bar{2}01)$ plane, and the incident electron beam is parallel to the [102] direction. The indentation load was 50 mN. A dislocation-rich region extending to a depth of ～1.6 μm is observed beneath the indentation. Dislocations labeled D-1 to D-8 indicate the dislocations analyzed in this study.

Fig. 2. Three kinds of bright-field LACBED patterns obtained from area D-1. The reciprocal lattice vector $\boldsymbol{g}$ and $n$ are (a) $\boldsymbol{g}_1 = \bar{2}\,\bar{2}\,1,\ n_1 = 2$, (b) $\boldsymbol{g}_2 = 7\,3\,\bar{3},\ n_2 = -3$, (c) $\boldsymbol{g}_3 = \bar{8}\,\bar{2}\,4,\ n_3 = -2$. The dashed line D, white arrow L, and number of yellow arrows represent the dislocation lines, Laue reflection lines, and number of nodes, respectively.

Fig. 3. WBDF images of dislocations formed around the indentation in β-$Ga_2O_3$. (a) With reciprocal lattice vector $\boldsymbol{g} = \bar{2}\ 0\ 1$ and (b) $\boldsymbol{g} = 0\ 2\ 0$. (c) and (d) are enlarged images of the same area indicated by the yellow rectangles. In (c), the contrast of dislocation D-1 is extremely weak, whereas in (d) it appears strong. The excitation condition for the dark-field images was $\boldsymbol{g}/3\boldsymbol{g}$ in both cases.

**Fig. 1** (rotated by 90°)

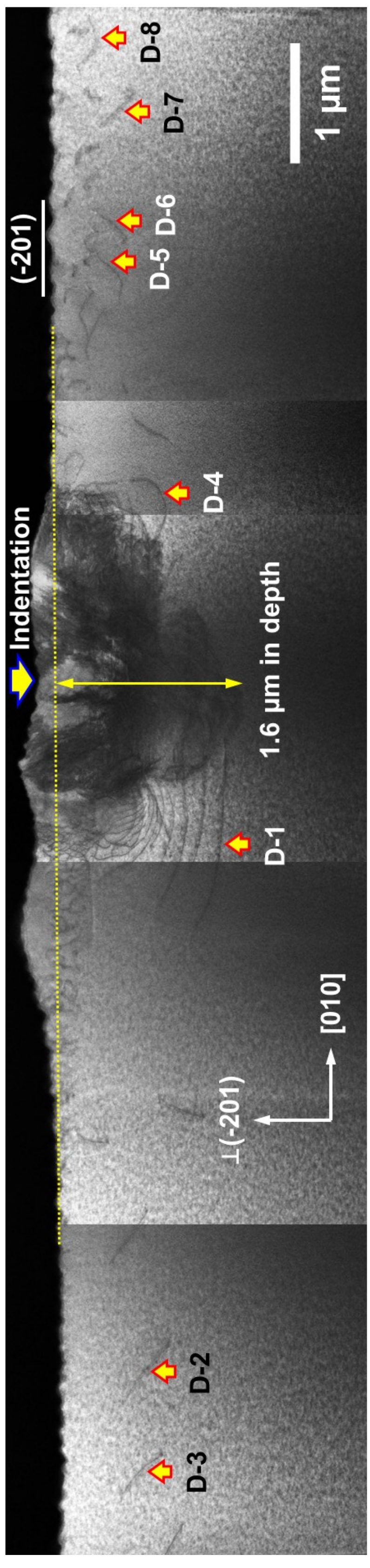

**Fig. 2** (rotated by 90°)

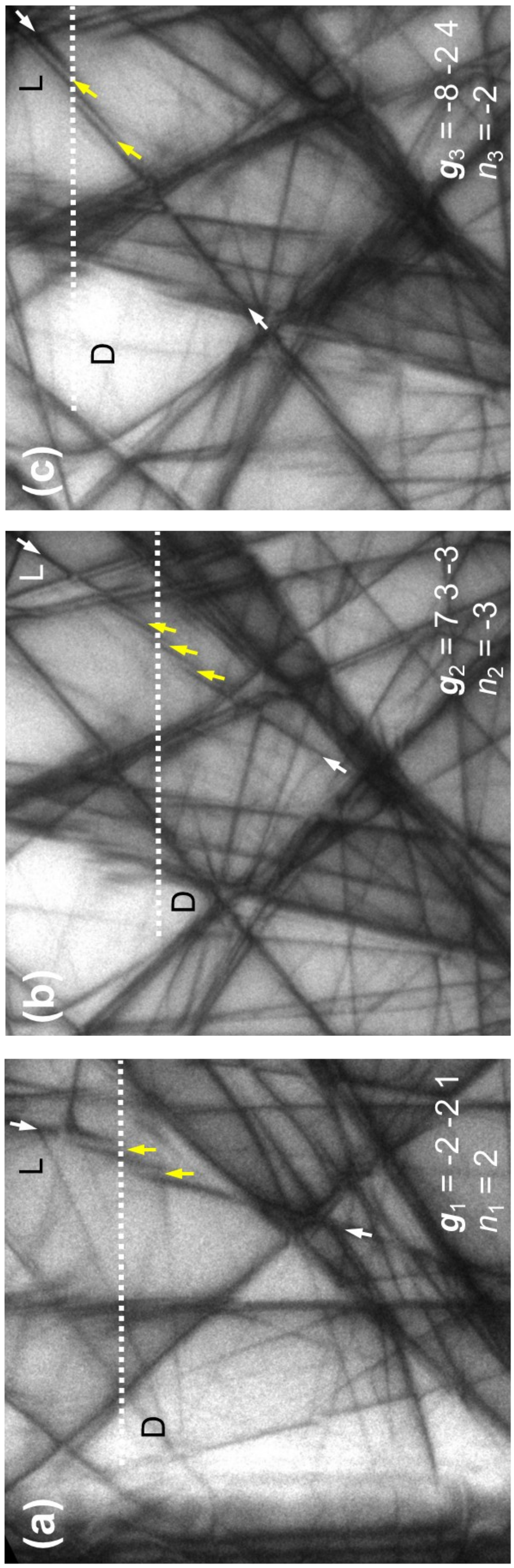

**Table 1**

| Dislocation No. | Reciprocal lattice vector, $\boldsymbol{g}_{hkl}$ | Node, $n$ | Burgers vector, $\boldsymbol{b}_{uvw}$ |
|---|---|---|---|
| D-1 | $\bar{2}\ \bar{2}\ 1$ | 2 | $[0\ \bar{1}\ 0]$ |
| | $7\ 3\ \bar{3}$ | -3 | |
| | $\bar{8}\ \bar{2}\ 4$ | 2 | |
| D-2 | $\bar{2}\ \bar{2}\ 1$ | -2 | $[0\ 1\ 0]$ |
| | $7\ 3\ \bar{3}$ | 3 | |
| | $\bar{8}\ \bar{2}\ 4$ | -2 | |
| D-3 | $\bar{2}\ \bar{2}\ 1$ | -2 | $[0\ 1\ 0]$ |
| | $7\ 3\ \bar{3}$ | 3 | |
| | $\bar{8}\ \bar{2}\ 4$ | -2 | |
| D-4 | $\bar{7}\ 3\ 4$ | 3 | $[0\ 1\ 0]$ |
| | $\bar{8}\ \bar{2}\ 4$ | -2 | |
| | $\overline{13}\ \bar{1}\ 7$ | -1 | |
| D-5 | $\bar{2}\ \bar{2}\ 1$ | -2 | $[0\ 1\ 0]$ |
| | $\bar{3}\ 3\ 2$ | 3 | |
| | $7\ \bar{3}\ \bar{4}$ | -3 | |
| D-6 | $\bar{1}\ 3\ 1$ | 3 | $[0\ 1\ 0]$ |
| | $\bar{3}\ 3\ 2$ | 3 | |
| | $\bar{7}\ 3\ 4$ | 3 | |
| D-7 | $7\ 3\ \bar{3}$ | -3 | $[0\ \bar{1}\ 0]$ |
| | $\bar{8}\ \bar{2}\ 4$ | 2 | |
| | $\overline{10}\ \bar{4}\ 5$ | 4 | |
| D-8 | $\bar{2}\ \bar{2}\ 1$ | 2 | $[0\ \bar{1}\ 0]$ |
| | $\bar{3}\ 3\ 2$ | -3 | |
| | $8\ 2\ \bar{4}$ | -2 | |

**Fig. 3** (rotated by 90°)

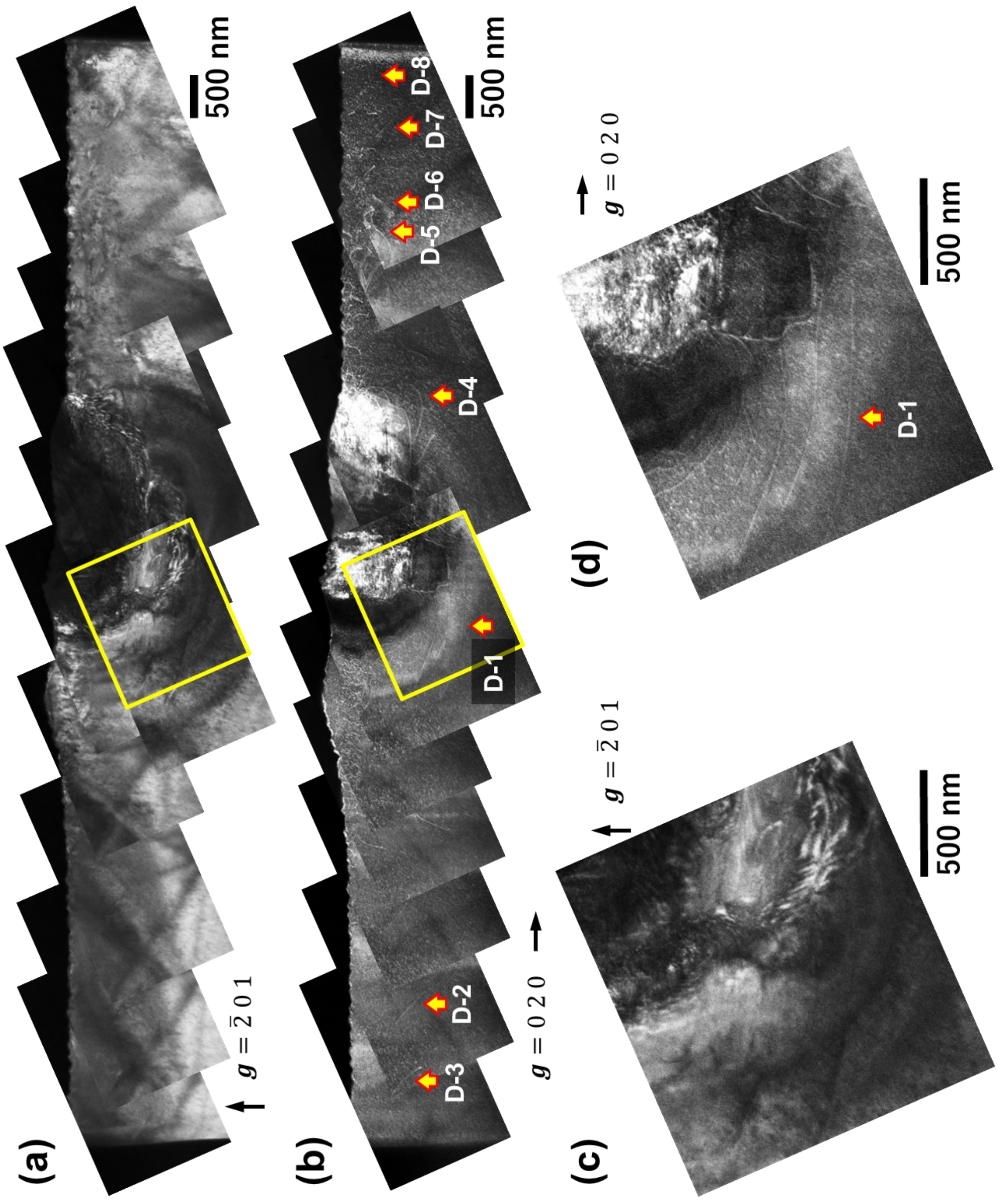